\documentclass[aps,prd,twocolumn,groupedaddress,nofootinbib,longbibliography]{revtex4-1}

\usepackage{amsfonts,amsmath,amssymb}
\usepackage{nicefrac}
\usepackage{graphicx}
\usepackage{color}

\usepackage{hyperref}

\usepackage[normalem]{ulem}

\def\e{\epsilon}
\def\o{\omega}
\def\w{\wedge}
\def\a{\alpha}
\def\b{\beta}
\def\k{\kappa}
\def\s{\sigma}

\def\a{\alpha}
\def\b{\beta}
\def\g{\gamma}
\def\D{\Delta}
\def\e{\epsilon}

\def\k{\kappa}

\def\o{\omega}
\def\O{\Omega}

\def\s{\sigma}

\def\del{\partial}

\def\half{{\textstyle{\frac{1}{2}}}}

\def\phib{{\bar\phi}}

\def\beq{\begin{equation}}
\def\eeq{\end{equation}}
\def\bea{\begin{eqnarray}}
\def\eea{\end{eqnarray}}

\begin{document}

\title{Blandford-Znajek process in vacuo and its holographic dual}

\author{Ted Jacobson}
\email[]{jacobson@umd.edu}
\affiliation{Center for Fundamental Physics, University of Maryland, 
                    College Park, Maryland 20742, USA}
\author{Maria J.~Rodriguez}
\email[]{maria.rodriguez@usu.edu\\maria.rodriguez@aei.mpg.de}

\affiliation{Max Planck for Gravitational Physics, Potsdam 14476, Germany}
\affiliation{Department of Physics, Utah State University, 4415 Old Main Hill Road, UT 84322, USA }

\begin{abstract}
Blandford and Znajek
discovered a process by which a spinning 
black hole can transfer rotational energy to a 
plasma, offering a mechanism for energy and 
jet emissions from quasars.
Here we describe a version of this mechanism that
operates with only vacuum electromagnetic fields
outside the black hole. 
The setting, which is not astrophysically realistic,
involves either a cylindrical black hole or one that 
lives in 2+1 spacetime dimensions,
and the field 
is given in simple, closed form for a wide class of metrics.
For asymptotically Anti-de Sitter black holes in 2+1 dimensions
the holographic dual of this mechanism is the transfer of 
angular momentum and energy, via a resistive coupling,
from a rotating thermal state containing an electric field to an 
external charge density rotating more slowly than the thermal state.
In particular, the entropy increase of the thermal state due to Joule heating
matches the Bekenstein-Hawking entropy increase of the black hole.

\end{abstract}


\maketitle

\section{Introduction}

Prodigious amounts of energy wind up in the 
rotation
of spinning black holes.
Although nothing can escape from inside a black hole, that energy can be extracted via a dynamo
effect, the {\it Blandford-Znajek (BZ) process} \cite{Blandford:1977ds}, 
when a black hole is threaded by a magnetic field and surrounded by plasma. 
The mechanism of the BZ effect is a type of {\it Penrose process} \cite{Penrose:1969pc}:
Near a spinning black hole is an {\it ergosphere} where it is impossible to remain stationary 
relative to a distant observer, because the local inertial frames are dragged around faster 
than the speed of light. A system in the ergosphere can carry negative values of the 
conserved global energy quantity, yet have positive ordinary energy relative to a local observer.
(In this paper the word ``energy" will refer to the globally conserved quantity, which agrees with 
``local energy" only far from the black hole, in its rest frame.)
Total energy can therefore be globally conserved, while positive energy flows away
and negative energy flows into the black hole.

Although the principles of the BZ process are fully understood  (e.g.\ \cite{Okamoto:2005hi,Lasota:2013kia}), 
and it is well studied analytically, semi-analytically, and numerically 
(e.g.\ \cite{Blandford:1977ds, Gralla:2015vta,Nathanail:2014aua,Komissarov2001,Komissarov:2007rc,McKinney:2006tf}),
it remains difficult to develop an intuitive picture of how the flow of electromagnetic energy and 
charge current is organized in space, and how and where the negative energy originates.
The difficulty is in part due
to the three dimensionality, the number of different field quantities, and the complexity of the Kerr metric 
which describes the spacetime geometry of a spinning black hole. In an effort to boil down the process to 
a ``toy model," we began looking for a version in 
a 2+1 dimensional spacetime.
We assumed that, as in the original BZ solution, 
the plasma is ``force-free", i.e.\ that it has a (nonzero) charge 4-current density orthogonal to the field strength,
but discovered that there is no such energy-extracting, stationary axisymmetric solution 
(see Appendix \ref{nogo} for a demonstration). 
Much to our surprise, however, we found that
there is a {\it purely electromagnetic} version of the BZ process,
in which plasma plays no role. 

We begin here with cylindrical BH examples, {since they are easier to visualize, and the analogy with the
astrophysical case is closer.} The 2+1 version will then be obtained as a special 
case of the 3+1 cylindrical one. Of particular interest is the spinning BTZ black hole background in 2+1 dimensions \cite{Banados:1992wn},
since it is an exact solution of general relativity with a negative cosmological constant, is locally maximally
symmetric, and is related by AdS/CFT duality  \cite{Maldacena:1997re} to a thermal state of a two dimensional conformal field theory. Thus one can also examine the CFT dual of a BZ process.
A similar idea was previously pursued in Ref.~ \cite{Wang:2014vza}, with force-free plasma in the Kerr-AdS background, 
but no energy extracting solution was found. It was suggested there that 
this might be due to the existence of a globally timelike Killing vector outside the event horizon, but this
implies only  that positive values of the corresponding conserved quantity cannot be extracted, and this need not be the relevant energy. Indeed, 
the spinning BTZ black hole also has a globally timelike Killing vector, yet we find that positive energy extraction is possible.

\section{Cylindrical BZ process}

\begin{figure}
\begin{center}
\includegraphics[width=6.0cm]{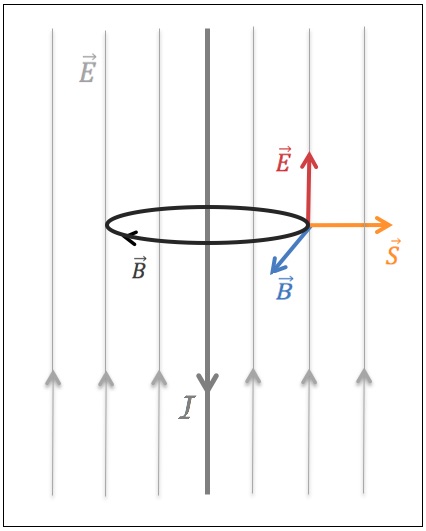}
\caption{\small {\it A line current driven opposite to an ambient electric field emits a Poynting flux of energy.
This field configuration corresponds to \eqref{simplesti}. Its generalization to a 
any stationary axisymmetric spacetime \eqref{ds2}, allowing for a radial component of the magnetic field, 
is given by \eqref{simplest}. For spinning black cylinder spacetimes, the line current is behind the horizon, and the field is nonsingular on the horizon provided the Znajek condition \eqref{Z} holds.
Energy is then extracted from the black cylinder if 
there is magnetic flux through the horizon and 
$0<\Omega_F<\Omega_H$ \eqref{JErH}.
The nonzero black hole spin is what  makes this extraction possible.}}
\label{line}
\end{center}
\end{figure}

In the BZ process \cite{Blandford:1977ds} a stationary plasma, 
around a spinning black hole threaded by a magnetic field, radiates a Poynting flux of energy to infinity. 
To simplify the geometry, 
and to eliminate the need for charge outside the black hole, 
we enhance the axisymmetry to  
cylindrical symmetry. Since the azimuthal circumference
grows as the cylindrical radius $r$, the radial component of the Poynting vector 
$\vec E\times\vec B$
must then 
fall as $1/r$
to conserve energy.
For example, if there is a uniform electric field parallel to the cylindrical 
axis, then the magnetic field must have an azimuthal, circular component that falls 
as $1/r$. 
Amp\`{e}re's law then implies that there must be an enclosed electric current parallel to the axis. 
In order for the Poynting flux to be radially outward, the flow of current must be opposite to the external electric field. The energy required to drive this current ends up in the outgoing Poynting flux. 
See Fig.~\ref{line} for an illustration of this scenario.

Consider this model first in flat, empty spacetime. 
If current flows only along the axis, the Poynting flux emerges from the axis.
Of course a line source of Poynting flux doesn't much resemble the BZ process, so let us introduce a cylindrical black hole, and hide the line source 
behind the event horizon.
A solution essentially like the one just described can be placed on this black hole spacetime but, since nothing can escape from a black hole, we expect that energy cannot be extracted from the source behind the horizon. Indeed, if we 
try to construct such a solution,
we find that the fields are singular on the horizon. In effect, the horizon 
becomes a singular source of energy, which again does not resemble the BZ process. 
However, if the black hole is spinning
with angular velocity $\O_H$, and if there is a nonzero magnetic flux through the horizon
that is related in a particular way to the electric field, the current, and $\O_H$,
then outgoing Poynting flux exists 
with fields that are regular on the horizon.
The electromagnetic dual of this solution 
has a uniform magnetic field parallel to the cylindrical axis, a radial electric flux through the horizon, and an azimuthal electric field. 
This reduces to a $2+1$ dimensional solution to Maxwell's equations when the 
direction along the axis is omitted. 
We turn now to a precise demonstration of these conclusions.

\subsection{Field Configurations}
We adopt cylindrical spacetime coordinates $(t,r,\phi,z)$, 
and assume the spacetime line element takes the form 
\beq\label{ds2}
ds^2 = -\a^2 dt^2 + \a^{-2}dr^2+ r^2(d\phi -\O_{} dt)^2  + dz^2, 
\eeq
where
 the functions $\a$ and $\O$ depend only on $r$, and
 we use units with $c=1$.
The spacetime is then
stationary and cylindrically symmetric: $ds^2$ is invariant
under translation of $t$, $\phi$, and $z$. We refer to the $t$ coordinate as ``time". 
This line element describes a wide class of spacetimes.
If there is a  
radius $r_H$ where $\a(r_H)=0$,
\eqref{ds2} describes a ``black cylinder" or ``black string,"  i.e.\ a black hole with a cylindrical horizon
at the outermost root $r_H$.
If $\O_H:=\O(r_H)\ne0$, the black cylinder is spinning with angular velocity $\O_H$,
and has an ergosphere where the time translation is spacelike, i.e.\
where $\a^2<r^2\O^2$.  
The BTZ black string metric \cite{Emparan:1999fd} is conformal to \eqref{ds2}. 
Since Maxwell's equations 
are conformally invariant, 
the solutions we discuss here are also solutions in the background of the BTZ black string.

{To describe the electromagnetic field we employ the language of differential forms,
which greatly simplifies computations, especially in curved spacetime.
(For a concise review of differential forms and their application to electromagnetism see Appendix A of \cite{Gralla:2014yja}.)
The electromagnetic field strength 2-form is denoted by $F$,  an index-free 
notation for the antisymmetric tensor $F_{ab}$. 
(For example, $F = E \,dx\w dt + B\, dx\w dy$ describes an electric field strength $E$ in the 
$x$ direction and magnetic field strength $B$ in the $z$ direction, in flat spacetime with Cartesian coordinates.)
Maxwell's equations are $dF=0$ 
(absence of magnetic monopoles and Faraday's law)
 and $d*F=J$ (Gauss and Amp\`{e}re-Maxwell laws),
where $d$ is the (metric independent) exterior derivative operator, 
$J$ is the current 3-form, and $*$ is the Hodge dual operation that sends a 
$p$ form $\o$ to the orthogonal $4-p$ form $*\o$ with the same magnitude. 
We adopt the orientation of $dt\w dr\w d\phi\w dz$ for defining the dual
({$\o\w*\o=\o^2\e$, where 
$\o^2$ is the squared norm of $\o$ and 
$\e$ is the unit 4-form with the adopted orientation}).
Note that the metric \eqref{ds2} is a sum of squares of 1-forms, hence 
those 1-forms are  orthonormal, which facilitates
the computation of duals and norms.

Let us begin with the simple example in flat spacetime, 
\beq\label{simplesti}
F =\left[\frac{I}{2\pi r} dr -E dt\right]\w dz,
\eeq
where $I$ and $E$ are constants. This describes a uniform electric field in the $\hat z$ direction, 
together with a magnetic field circulating around the origin in planes of constant $t$ and $z$, 
sourced by a line current at $r=0$ flowing in the $-\hat z$ direction. 
The dual of the field \eqref{simplesti} is
\beq\label{*simplesti}
* F= \frac{I}{2\pi}d\phi\w dt -Erd\phi\w dr.
\eeq
It is clear by inspection that Maxwell's equations $dF=0=d*F$
are satisfied (since $d^2=0$ and $dr\w dr=0$), so this is indeed a vacuum solution, except 
at the axis where $d\phi$ is singular.
The field \eqref{simplesti} is the wedge product of two 1-forms, 
which implies that $F\w F=0$. 
Equivalently, the Lorentz-invariant scalar $\vec E\cdot\vec B$ vanishes, 
as it does for ideal plasmas since $\vec E$ vanishes in the 
rest frame of a perfect conductor. A field with this algebraic 
property is called {\it degenerate}.

The field \eqref{simplesti} 
can be easily generalized to the case of any background 
spacetime with a metric of the cylindrical form \eqref{ds2}.
Allowing also for a radial component of the magnetic field, 
the result is given by 
\beq\label{simplest}
F =\left[\frac{I}{2\pi r\a^2} dr+ \psi_{z}(d\phi -\O_F dt)\right]\w dz,
\eeq
where $I$, $\psi_{z}$, and $\Omega_F$ are constants.
The dual of \eqref{simplest} is
\beq\label{*simplest}
* F= \frac{I}{2\pi}d\phi\w dt +\psi_z\left[\frac{1}{r} dt + \frac{r(\O-\O_{F})}{\a^2}(d\phi-\O_{} dt)\right]\w dr,
\eeq
As with \eqref{simplesti}, Maxwell's equations are satisfied by inspection. 
(The functions $\a$ and $\O$ depend only on $r$, so that $d\a\w dr =0=d\O\w dr$.)
The field \eqref{simplest} remains degenerate, since it is the wedge product of two 1-forms.
To recover \eqref{simplesti} from \eqref{simplest} one should set 
$\alpha=1$ and send $\psi_z$ to zero and $\Omega_F$ to infinity, 
with $E:=\psi_z\Omega_F$ held fixed.
The constant parameters (and their notation) in \eqref{simplest}
are directly analogous to functions that appear in standard treatments
of stationary axisymmetric plasma magnetospheres (e.g.\ \cite{MacDonald:1982zz,Gralla:2014yja}
and references therein). \\

The singularity on the axis carries a line current $I$
in the $-z$ direction,
sourcing the azimuthal magnetic field, 
and a magnetic monopole line charge density $2\pi \psi_{z}$
sourcing the radial magnetic field. 
($2\pi \psi_z$ is the $z$-derivative of the magnetic flux
through a cylinder ending at coordinate $z$.)
The quantity $\O_F$ is the ``angular velocity of 
the magnetic field lines," in the sense that the electric field vanishes 
in the local frame that rotates that way: $(\del_t+\O_F\del_\phi)\cdot F=0$. This is
a standard concept in ideal plasma physics, where the plasma determines the
preferred frame. It is meaningful here without the plasma only
because the field is degenerate.

We are interested in the solution \eqref{simplest} on a rotating black hole
background, but it is worth noting that it could also be terminated on 
a cylindrical conductor rotating with angular velocity $\O_F$.

\subsection{Energy Extraction}
\label{IIB}
The Maxwell action is given by $-\half\int F\w *F$, from which it follows that 
the conserved Noether current associated with the $t$-translation symmetry 
of the spacetime is \cite{Gralla:2014yja}
\beq\label{JE}
{\cal J}_E = -(\partial_t\cdot F)\w *F + \half \partial_t\cdot(F \w * F).
\eeq
As is usual, we call this conserved quantity simply ``energy". 
The outgoing energy flux is given by the integral of {the 3-form} ${\cal J}_E$ over a surface of constant $r$, 
which by virtue of the conservation law is the same as the flux across any other surface of constant $r$. 
Replacing $\partial_t \rightarrow -\partial_{\phi}$ in (\ref{JE}), one obtains the 
angular momentum current ${\cal J}_L$ associated with the $\phi$-translation 
symmetry.

The energy current outward through a surface of constant $r$ 
arises from the part of ${\cal J}_E$ that contains no $dr$ factor (so the 
second term on the right hand side of \eqref{JE} does not contribute), 
which for the field \eqref{simplest} is given by 
\beq\label{JEr}
{\cal J}_E|_r =\frac{1}{2\pi}{\O_F\psi_{z} I}\, dt\w d\phi\w dz.
\eeq
The outward flux of this current over a constant $r$ cylinder of length 
$\D z$ and time interval $\D t$ is ${\O_F\psi_{z} I} \D t \D z$, so the energy and angular momentum
fluxes per unit proper length per unit $t$-coordinate time at any radius are given by
\beq\label{PhiE}
{\cal E}_r=\O_F\psi_{z} I,\quad {\cal L}_r=\psi_{z} I.
\eeq
The outgoing energy flux 
is positive if $I$ has the
same sign as $\O_F\psi_{z}$, which means that the current is opposite to the electric field. 
This corresponds to the situation described in the introduction to this section, 
where the line current driven against
the electric field produces an outgoing Poynting flux. 

The line element \eqref{ds2} indicates that 
$\a\, dt$ and $dr/\a$ are unit 1-forms, so that $dt$ and $dr/\a^2$ are both singular at $\a=0$
in a black cylinder spacetime.
Therefore the $I$ and $\O_F$ terms in the field \eqref{simplest} are both singular at the horizon. 
However,  the forms $dt$ and 
$dr$ become proportional on the horizon, and their divergences can cancel. 
To analyze this divergence balancing act, one can 
either transform to a new coordinate system that is 
regular on the horizon, or construct scalars by contracting
$F$ with a basis of finite vectors. One finds (see Appendix \ref{Zderivation}) that
a necessary condition for finiteness at the future horizon is 
\beq\label{Z}
I=2\pi r_H\psi_{z}(\O_H-\O_F).
\eeq
This condition is also sufficient, provided 
$\a^2$
has nonvanishing first derivative at $r_H$
(i.e., provided the black hole is not extremal). 
The relation \eqref{Z} is known as the {\it Znajek condition} \cite{Znajek:1977}. 

If the regularity condition \eqref{Z}
holds, then the outgoing energy flux \eqref{PhiE} becomes
\beq\label{JErH}
{\cal E}_{r}=  2\pi (\psi_{z})^{2}\,r_H\O_F(\O_H-\O_F).
\eeq
If the black cylinder is not rotating ($\O_H=0$), then the outward energy flux 
is always negative. This means that, as expected, 
while energy can flow inward
it cannot be extracted. However,
if the black cylinder is rotating, and if $0<\O_F<\O_H$, then 
positive energy can be radiated outward 
provided there is a nonzero magnetic flux $\psi_z$ through the horizon. 
The energy flux formula \eqref{JErH} is precisely analogous to that for 
a force-free plasma on a Kerr black hole spacetime \cite{Blandford:1977ds,MacDonald:1982zz,Gralla:2014yja}.

The vacuum Maxwell equations are invariant under electric-magnetic duality, $F\rightarrow *F$,
as is the energy momentum tensor and Noether current \eqref{JE}, so the dual \eqref{*simplest} of the field \eqref{simplest} provides another solution that extracts energy from the black cylinder. The dual solution has a magnetic field in the $z$ direction, and azimuthal and radial electric field components. 
The radial electric field is sourced by an electric line charge, and 
the azimuthal electric field can be sourced by either a magnetic monopole current or a magnetic flux line with flux proportional to time. 

\section{$2+1$-Dimensional Model}

The dual field strength \eqref{*simplest} has no $dz$ factor, and is invariant under $z$ translations, so it descends to a solution 
$F_3:=*_4 F$ in three-dimensional spacetime. The three-dimensional Hodge dual $*_3F_3$ 
is a 1-form, which is just the negative of the first factor of our original solution, \eqref{simplest}.
In the remainder of this article, we shall focus on this example. Maintaining the convention that the 
vector potential is an inverse length, in three spacetime dimensions the Maxwell action 
written at the beginning of Sec.~\ref{IIB}
should be multiplied by 
a constant with dimensions of length, $1/g^2$. We adopt units here with $g=1$.

Dualizing interchanges electric and magnetic quantities, so we change our notation for the parameters accordingly: 
$Q:=-2\pi\psi_z$ is the electric charge of the solution, and 
$\dot\Phi:=-I$ is the $t$ derivative of the magnetic flux through a loop encircling the origin (or the magnetic monopole current).
The constant $\O_F$ is now the ``angular velocity of the electric field lines", that is, the 
angular velocity of the frame in which the magnetic field (which in 2+1 dimensions is a spatial pseudoscalar) 
vanishes (presuming the vector $\partial_t +\O_F\partial_\phi$ is timelike,
which is here the case for large enough $r$).
The three dimensional field and its dual are given by\footnote{All stationary, axisymmetric vacuum solutions to Maxwell's equations in a 2+1 dimensional spacetime (with the same symmetry) have this form.}
\begin{align}
F_3&=-\frac{\dot\Phi}{2\pi}\,d\phi\w dt -\frac{Q}{2\pi r} dt\w dr\nonumber\\
&~~~~-\frac{Qr(\O-\O_{F})}{2\pi \a^2}(d\phi-\O_{} dt)\w dr,\label{F3}\\
*_3\,F_3 &=\frac{\dot\Phi}{2\pi\alpha^2r}\,dr+\frac{Q}{2\pi}(d\phi- \Omega_F\, dt).
\label{eq:F3dual}
\end{align}
The Znajek horizon regularity condition \eqref{Z} 
in the present notation is $\dot\Phi = Qr_+(\O_H-\O_F)$.
The dual of the regular field is thus 
\beq
*_3F_3 =\frac{Q}{2\pi}\left[\frac{r_+(\O_H-\O_F)}{\alpha^2r}\,dr+d\phi- \Omega_F\, dt\right].
\label{eq:F3dualreg}
\eeq

This solution can be placed in particular on the rotating BTZ black hole background \cite{Banados:1992wn}, 
which is a three-dimensional solution of Einstein's equations
with a negative cosmological constant, $\Lambda_3=-\ell^{-2}$.
The corresponding line element (in Boyer-Lindquist-like coordinates) is (\ref{ds2}) without $dz^2$, 
and with
\beq\label{BTZfunctions}
\alpha^2=\frac{(r^2-r^2_+)(r^2-r^2_-)}{r^2\ell^2},\qquad \Omega=\frac{r_-r_+}{r^2\ell},\,
\eeq
where $r_-$ and $r_+$ are the inner and outer horizon radii, respectively, with $0<r_-<r_+$. 
On this spacetime, the invariant square of the field strength is 
\beq \label{electric}
F_3^2 = \frac{Q^2}{4\pi^2}\frac{\ell^2\O_F^2-1}{r^2 - r_-^2}.
\eeq
In the energy extracting case ($0<\O_F<\O_H$) we have $\ell\O_F<\ell\O_H\le1$,
so this field is electric dominated everywhere 
outside the inner horizon, and is singular at the inner horizon. (For the case $\O_F=\O_H$ this singularity
has been previously noted \cite{Martinez:1999qi}.)
In Appendix \ref{C} we express the BTZ metric and the electromagnetic field
in terms of ingoing Kerr coordinates $(v,r,\phib)$, which are regular 
everywhere except at $r=0$. The electric field lines on a surface of
constant $v$ are plotted in Fig. \ref{ergo}.
\begin{figure}[h!]
\begin{center}
\includegraphics[width=7cm]{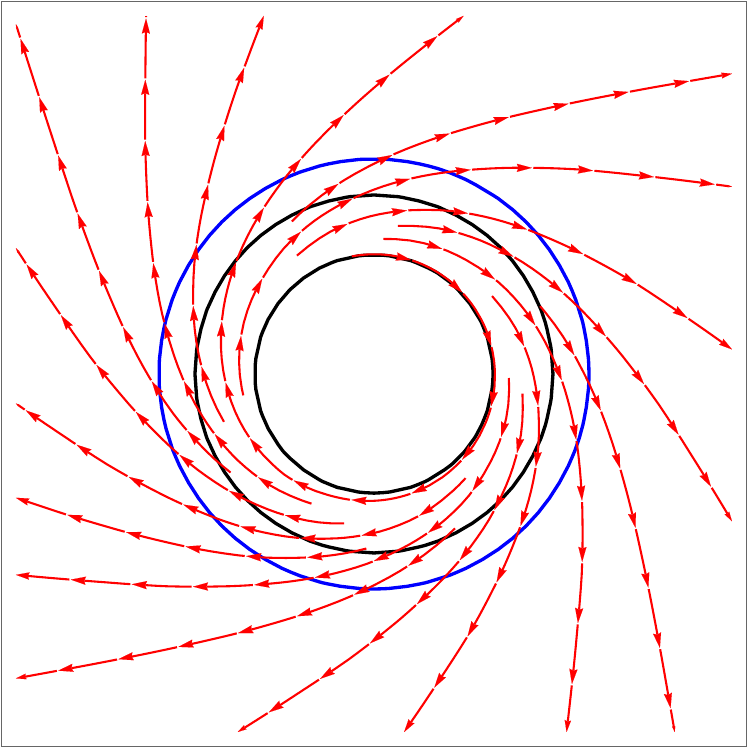}
\caption{\small {\it Electric field lines for the field \ref{eq:F3dualreg}, with $r_+=0.3$, $r_-=0.2$, and $\O_F=\O_H/2 = 1/3$, in units with $\ell=1$.
As explained in Appendix \ref{C}, the field lines are intersections of the electric field sheets with a surface of constant ingoing Kerr coordinate $v$,  plotted with $r$ and $\phib$ treated as polar coordinates on the Euclidean plane. The field lines for $r<r_-$ are not plotted, since $F_3$ is singular at $r_-$.}}
\label{ergo}
\end{center}
\end{figure}

\subsection{CFT dual}
We now turn to the 
AdS/CFT dual interpretation
of the vacuum BZ process on the spinning BTZ black hole background.
A number of previous studies have employed
this duality with Einstein-Maxwell theory in three spacetime dimensions, sometimes with other fields included (see e.g.\  \cite{Maity:2009zz,Jensen:2010em,Ren:2010ha,Faulkner:2012gt,Horowitz:2013mia,Chaturvedi:2013ova}), 
to study conductivity properties of the dual 1+1 dimensional system.
The black hole is dual to a thermal state of a CFT at the Hawking temperature, 
rotating with angular velocity $\O_H$ \cite{Hawking:1998kw}. To extract rotational energy from 
this thermal state, it must be coupled to another system, and that coupling is described 
by the boundary conditions on the 
electromagnetic field. In the limit of large radius $r$, the field $F_3$ \eqref{F3}
can be derived (as $F_3=dA_3$) from the potential 
\bea
A_3\xrightarrow{ r \to \infty } \frac{\dot\Phi}{2\pi} t\,d\phi-\frac{Q}{2\pi} \ln r\, (dt-\ell^2\Omega_F \,d\phi),
\eea
which corresponds, via the AdS/CFT dictionary \cite{Marolf:2006nd,Jensen:2010em}, to a boundary 
CFT current $j$ and gauge field $a$,
\bea\label{j&a}
j=\frac{Q}{2\pi\ell}(\del_t + \Omega_F\del_\phi),\qquad   a=\frac{\dot\Phi}{2\pi}\,t\,d\phi.
\eea
One dynamically consistent boundary condition
fixes $j$ \cite{Marolf:2006nd}, which fixes the black hole charge and the asymptotic magnetic field.
The transfer of energy between the CFT and the (unspecified) degrees of freedom carrying the current then entails Ohmic dissipation. 

The conductivity $\s$ of the CFT is the ratio of the current to the electric field, 
figured in the rest frame of the thermal state.
That is, $\s=(s\cdot j)/(u\cdot s\cdot f)$, where $f=da=(\dot\Phi/2\pi)dt\w d\phi$ is the boundary electromagnetic field strength, $u= \g(\partial_t + \O_H\partial_\phi)$ is the velocity 2-vector of this rotating frame, and $s=\g(\ell\O_H\partial_t +\ell^{-1}\partial_\phi)$ is the orthogonal, spatial unit 2-vector (with respect to the boundary metric, $ds^2 = -dt^2 + \ell^2 d\phi^2$), 
with $\g=(1-\ell^2\O_H^2)^{-1/2}$.
After using the Znajek horizon regularity condition, 
the dependence on the 
electromagnetic field parameters drops out, and we find
\beq\label{sigma}
\s = \frac{\g\ell}{r_+}= \frac{1}{\g\k\ell}=\frac{\hbar}{2\pi\gamma T_H},
\eeq
where $\kappa$ is the surface gravity of the BTZ black hole, and $T_H=\hbar\kappa/2\pi$ 
is the Hawking temperature. In the nonrotating limit, 
$\sigma$  is inversely proportional to the Hawking temperature, in agreement
a prior result \cite{Horowitz:2013mia}. 
The conductivity depends on angular momentum of the black hole only via the Lorentz boost factor $\g$,
and $\g T_H$ is the temperature in the corotating frame of the thermal state \cite{Frolov:1989jh}, so in fact
the rotating conductivity is the same as in the nonrotating case when expressed in terms of the corotating temperature.

\subsection{Black Hole Evolution}
The response 
to weak fluxes of electromagnetic energy and angular momentum
can be determined using 
the laws of black hole mechanics (which are derived from Einstein's equation).
The flux formula \eqref{PhiE} shows that $dE = \O_F dL$ for the electromagnetic fluxes of energy and angular momentum, 
and conservation laws imply that infinitesimal changes of the black hole mass $M$ and angular momentum $J$ are related 
in the same way, $dM = \O_F dJ$. The First Law of black hole mechanics \cite{Bardeen:1973gs},
gives $dM -\O_H dJ = (\kappa/8\pi G)dA$, where $\kappa$ and $A$ are the surface gravity and area of the horizon, respectively,
hence we have $(\O_F-\O_H) dJ = (\kappa/8\pi G)dA$. The Second Law of black hole mechanics \cite{Hawking:1971vc} 
states that $dA\ge0$, in processes with matter satisfying the null positive energy condition (such as electromagnetic fields). 
Thus, if $\O_F<\O_H$, the black hole spins down, and loses or gains mass according as $\O_F$ is positive or negative. If instead 
$\O_F>\O_H$, the black hole spins up and gains mass.
If the asymptotic electric and magnetic fields are held fixed, then $\O_F$ is fixed, and
$\O_H$ evolves until it exponentially approaches the condition $\O_H=\O_F$, for which the fluxes vanish.

The black hole entropy 
$S_{\rm BH}=A/4\hbar G$ evolves as 
\beq\label{Sdot}
\dot{S}_{\rm BH} = \frac{\dot{M}-\O_H\dot{J}}{T_H}=\frac{(\O_F-\O_H)\dot{J}}{T_H}=\frac{\dot\Phi Q(\O_H-\O_F)}{2\pi T_H},
\eeq
where the overdot signifies derivative with respect to the Killing time $t$, and $T_H=\hbar\kappa/2\pi$ is the Hawking temperature. [The first equality invokes the first law of black hole mechanics, while the second and third make use of \eqref{PhiE} with the notational change indicated in the paragraph containing \eqref{F3}.]

The rate of entropy generation per unit spacetime volume, due to Ohmic dissipation in the holographic dual field theory process, is given by the rate of energy dissipation divided by the temperature in the corotating frame of the thermal state. The energy dissipation rate is the inner product of the electric field with the charge current in that frame, hence the total rate of entropy generation is  
\beq
\dot S_{\rm FT}=\frac{(u\cdot j\cdot F)2\pi \ell}{\gamma T_H}.
\eeq
This agrees with \eqref{Sdot} when the previously given expressions for $u$, $j$ and $F$ are inserted.
(This is special case of a very general relation between boundary and black hole entropy generation that has been established in \cite{Tian:2014goa}.)
It is amusing to note that this dual description is a perfect electrical analogy for the mechanical analogy
originally described by Blandford and Znajek  \cite{Blandford:1977ds}. That analogy 
involved a thermally conducting disk spinning with angular velocity $\O_H$,  coupled by friction to a 
concentric, thermally insulating ring spinning with a smaller angular velocity.

\section{Discussion}
We have thus far treated the electromagnetic field as a test field on the background of a spinning BTZ black hole background, and inferred adiabatic evolution from the conservation laws.
To more fully probe the duality one should solve the coupled Einstein-Maxwell equations. A similar study was carried out in \cite{Horowitz:2013mia} for the case of 
of electromagnetic energy flowing into a nonrotating black hole,
allowing the nonlinear conductivity of the dual field theory to be found.
In our case, a first step would be to use for the background the charged, spinning black hole solution  \cite{Clement:1995zt,Martinez:1999qi}, which also has a magnetic field with $\O_F=\O_H$ in our notation. This is the solution to which the black hole would relax,
 and the azimuthal electric field alone could be treated as a test field.  
The next step would be to attempt to find an Einstein-Maxwell solution including the azimuthal electric field, which could describe 
nonlinear effects in the conductivity. Another avenue to investigate would be alternate boundary conditions at infinity,
which can modify the nature of the dual field theory degrees of freedom and the external source that couples to them \cite{Marolf:2006nd,Jensen:2010em,Faulkner:2012gt}.

\begin{acknowledgments}
We thank S.~Hartnoll, G.~Horowitz, N.~Iqbal, K.~Jensen, D.~Marolf, A.~Speranza and S.~Theisen for helpful discussions and correspondence.
This research was initiated at the Peyresq Physics 21 meeting, and partially supported there by 
OLAM.  TJ was also 
supported in part by NSF grants PHY-1407744 and PHY-1708139 at the University of Maryland, 
and by Perimeter Institute for Theoretical Physics. Research at Perimeter Institute is supported by the Government
of Canada through Industry Canada and by the Province of Ontario
through the Ministry of Research and Innovation.
The work of MJR was supported by the Max Planck Gesellschaft through the Gravitation and Black Hole Theory Independent Research Group and by NSF grant PHY-1707571 at Utah State University. 
\end{acknowledgments}

\appendix 
\section{Stationary axisymmetric force-free solutions in 2+1 dimensions}
\label{nogo}

In three spacetime dimensions, the force-free condition $F_{ab}j^a=0$
is equivalent to the condition 
\beq\label{FF2}
d*F=\zeta F,
\eeq
for some function $\zeta$.
If $F$ is stationary and axisymmetric then $*F$ has the form 
$*F=\a(r) dt + \b(r) d\phi + \g(r) dr$, so $d*F = \a' dr\w dt + \b' dr \w d\phi$.
The force-free condition \eqref{FF2} thus implies that either the charge
current vanishes,  or $F\propto \a' dr\w dt + \b' dr \w d\phi$, in which case 
$F$ has no $dt\w d\phi$ term (i.e.\ no azimuthal electric field). 
But this implies that the energy current
\beq\label{JEE}
{\cal J}_E = -(\partial_t\cdot F)\w *F + \half \partial_t\cdot(F \w * F),
\eeq
 has no  $dt\w d\phi$ term,
so there is no radial energy flux. (Equivalently, the Poynting vector has no radial
component.)

\section{Derivation of the Znajek condition}
\label{Zderivation}

To characterize the regularity condition for the field strength 
\beq\label{simplestt}
F =\left[\frac{I}{2\pi r\a^2} dr+ \psi_{z}(d\phi -\O_F dt)\right]\w dz,
\eeq
on the horizon, one can 
either transform to a coordinate system that is 
regular on the horizon, or characterize $F$ by the 
scalars obtained when contracting
it with a basis of regular vectors. Here we use the second method.
For three of the four basis vectors we can use the Killing vectors, 
$\del_t$, $\del_\phi$, and $\del_z$,
which are all tangent to the horizon. 
(Note that although the coordinates $t$ and $\phi$ are singular on the horizon
of a spinning black hole, the vectors $\del_t$ and $\del_\phi$ are the time translation 
and rotation Killing vectors and, regardless of what coordinate system is used, 
these vectors are regular on the horizon.) 
The contractions of the three Killing vectors with $F$ are finite 
as long as the constants $\psi_{z}$ and $\O_F$ are finite.
For the fourth regular basis vector we
use the 4-velocity $k$ of  null ($k\cdot k=0$), zero angular momentum ($k\cdot\del_\phi=0$) 
and zero longitudinal momentum ($k\cdot\del_z=0$)
geodesics, with affine parameter fixed by the choice $k\cdot \del_t=-1$. 
These conditions (which are conserved along the geodesic) 
determine $k$ uniquely up to the choice of whether the geodesic is ingoing or outgoing. 
The {\it ingoing} one is
\beq\label{u}
k = \a^{-2}(\partial_t +\O_{}\partial_\phi) - \partial_r.
\eeq
The additional nonzero scalar that can be formed using $k$ is 
\beq\label{finite}
k\cdot\del_z\cdot F=\frac{\psi_{z}(\O-\O_F) - {I}/{2\pi r}}{\a^2}.
\eeq
\\
A necessary condition for finiteness at the future horizon
is therefore
\beq\label{ZZ}
I=2\pi r_H\psi_{z}(\O_H-\O_F).
\eeq
This is also sufficient, provided $(\a^2)_{,r}$ is nonvanishing at $r_H$, 
as is the case as long as the black hole is not extremal.

\section{Ingoing Kerr coordinates for the BTZ black hole}
\label{C}

The spinning BTZ black hole metric is given in Boyer-Lindquist coordinates
by \eqref{ds2} less the $dz^2$ term, and with $\a$ and $\O$ given by 
\eqref{BTZfunctions}. The coordinates $t$ and $\phi$ are singular at the horizon in this 
spacetime. In their place, we can 
introduce new coordinates $v= t+ f(r)$, 
and $\phib =\phi + g(r)$,
which will be regular if 
the functions $f(r)$ and $g(r)$ are defined
so that $v$ and 
$\phib$ are constant 
on the infalling, zero angular momentum null geodesics, whose affine 4-velocity $k$ is given by \eqref{u}.  The requirements $k\cdot dv = 0$ and $k\cdot d\phi=0$
then impose 
$f'=\a^{-2}-1$ and $g'=\a^{-2}$ respectively, hence we have 
\beq
dv = dt + \frac{1}{\a^2}\, dr,\qquad d\phib = d\phi + \frac{\O_{}}{\a^2}\, dr.
\eeq
In terms of these regular differentials, 
the metric takes the form
\beq\label{BTZ2}
ds^2 = -\a^2 dv^2 + 2\, dv \,dr+ r^2(d\phib -\O_{} dv)^2,
\eeq
which is regular except at $r=0$. This is called the ingoing Kerr coordinate system.

When expressed using these regular coordinates,
the dual field \eqref{eq:F3dual} takes the form
\beq\label{F3Kerr}
*_3 F_3 = \frac{Q}{2\pi}\left[\frac{\ell(r_- +\ell \Omega_Fr)}{(r+r_+)(r^2-r_-^2)}\,dr +d\bar{\phi} - \Omega_F\, dv\right].
\eeq
(We write the dual 1-form since it is simpler than the field strength 2-form $F_3$.) Since there is no globally defined, natural time coordinate relative to which one might define the electric field vector, we capture 
the `shape' of the field \eqref{F3Kerr} as follows. 
The kernel of $*_3 F_3$ at each point is two-dimensional, and is timelike outside the inner horizon since 
$F_3$ is electric dominated there \eqref{electric}. 
Since there is no charge current, $*_3 F_3$ is closed ($d*_3 F_3=0$), so this distribution of  planes is integrable, i.e.\ they are tangent to two-dimensional submanifolds, which can be called the ``electric field sheets" (see \cite{Gralla:2014yja} for a discussion of the magnetic case in four spacetime dimensions). 
The intersection of an electric field sheet with a spacelike surface is an electric field line on that surface. 
However, since we do not have preferred spacelike surfaces in the BTZ spacetime, we employ the intersection
with a constant $v$ null surface to generate a picture of the electric field lines. They are tangent to a vector
field $X$ satisfying $X\cdot*_3 F_3=0= X\cdot dv$. One more condition is needed to specify the magnitude of $X$, although this magnitude will not affect the field lines. If we choose for this condition $X\cdot dr=1$, 
then 
\beq
X = \partial_r -\frac{\ell(r_- +\ell \Omega_Fr)}{(r+r_+)(r^2-r_-^2)} \partial_{\phib}, 
\eeq
In Fig. \ref{ergo} we plot the stream lines of this vector field, treating $r$ and $\phib$ as polar coordinates
on the Euclidean plane.

\bibliography{BZwithoutPlasma}

\end{document}